# Titanium doped kagome superconductor $CsV_{3-x}Ti_xSb_5$ and two distinct phases

*Haitao Yang[1,2,3#], Zihao Huang[1,2#], Yuhang Zhang[1,2#], Zhen Zhao[1,2#], Jinan Shi[2], Hailan Luo[1,2], Lin Zhao[1,2], Guojian Qian[1,2], Hengxin Tan[4], Bin Hu[1,2], Ke Zhu[1,2], Zouyouwei Lu[1,2], Hua Zhang[1,2], Jianping Sun[1,2], Jinguang, Cheng[1,2], Chengmin Shen[1,2], Xiao Lin[2], Binghai Yan[4], Xingjiang Zhou[1,2], Ziqiang Wang[5], Stephen J. Pennycook[2,1], Hui Chen[1,2,3\*], Xiaoli Dong[1,2\*], Wu Zhou[2,3\*], and Hong-Jun Gao[1,2,3\**

[1] Beijing National Center for Condensed Matter Physics and Institute of Physics, Chinese Academy of Sciences, Beijing 100190, PR China

[2] School of Physical Sciences, University of Chinese Academy of Sciences, Beijing 100190, PR China

[3] CAS Center for Excellence in Topological Quantum Computation, University of Chinese Academy of Sciences, Beijing 100190, PR China

[4]Department of Condensed Matter Physics, Weizmann Institute of Science, Rehovot 7610001, Israel

[5] Department of Physics, Boston College, Chestnut Hill, MA, 02467, USA

[#]These authors contributed equally to this work

[*]Correspondence to: hjgao@iphy.ac.cn; wuzhou@ucas.ac.cn; dong@iphy.ac.cn; hchenn04@iphy.ac.cn

ABSTACT

The vanadium-based kagome superconductor $CsV_3Sb_5$ has attracted tremendous attention due to its unexcepted anomalous Hall effect (AHE), charge density waves (CDWs), nematicity, and a pseudogap pair density wave (PDW) coexisting with unconventional strong-coupling superconductivity (SC). The origins of CDWs, unconventional SC, and their correlation with different electronic states in this kagome system are of great significance, but so far, are still under debate. Chemical doping in the kagome layer provides one of the most direct ways to reveal the intrinsic physics, but remains unexplored. Here, we report, for the first time, the synthesis of Ti-substituted $CsV_3Sb_5$ single crystals and its rich phase diagram mapping the evolution of intertwining electronic states. The Ti atoms directly substitute for V in the kagome layers. $CsV_{3-x}Ti_xSb_5$ shows two distinct SC phases upon substitution. The Ti slightly-substituted phase displays an unconventional V-shaped SC gap, coexisting with weakening CDW, PDW, AHE, and nematicity. The Ti highly-substituted phase has a U-shaped SC gap concomitant with a short-range rotation symmetry breaking CDW, while long-range CDW, twofold symmetry of in-plane resistivity, AHE, and PDW are absent. Furthermore, we also demonstrate the chemical substitution of V atoms with other elements such as Cr and Nb, showing a different modulation on the SC phase and CDWs. These findings open up a way to synthesise a new family of doped $CsV_3Sb_5$ materials, and further representing a new platform for tuning the different correlated electronic states and superconducting pairing in kagome superconductors.

## 1. Introduction

The newly discovered kagome metals $AV_3Sb_5$ (A=K, Rb, Cs) have attracted tremendous research interest as a novel platform to study the interplay between nontrivial band topology, superconductivity (SC), and multiple density waves[1-7]. The $AV_3Sb_5$ carries no magnetic order at low temperature but exhibits an unconventional anomalous Hall effect (AHE) that is usually observed in ferromagnetic materials[8-10]. Most recently, unidirectional charge order[11, 12], electronic nematicity[13] and roton pair density wave (PDW)[6] have been observed in the strong-coupling kagome superconductor $CsV_3Sb_5$. Time-reversal symmetry breaking (TRSB) was proposed to originate from the unconventional charge density waves (CDWs)[2, 14], but this is still under debate[15-22]. The nature of the CDWs and superconducting state remains elusive. Theoretically, $CsV_3Sb_5$ has a $Z_2$ nontrivial topological band structure with topological surface states, indicating an unconventional superconductor[1, 23], while some theoretical works propose the conventional s-wave SC in $CsV_3Sb_5$. Experimentally, transport evidence for gap nodes[24] and gapless core states in magnetic-field induced vortices[3] have been reported, while tunneling diode oscillator[25] and nuclear quadrupole resonance[26] experiments showed the nodeless superconductivity nature of $CsV_3Sb_5$. Therefore, full understanding of the pairing of the superconducting state and its correlation with the kagome geometry, TRSB, and rotational symmetry breaking orders remains an important issue to be urgently resolved.

Chemical doping and external pressure both provide effective ways to study the nature of the superconducting state and its correlation with the other orders. For example, double superconducting domes with enhanced SC transition temperature have recently been observed in $CsV_3Sb_5$ upon applied pressure[5, 27]. However, it is difficult to combine the high-pressure approaches with the local electronic states and band structures, which makes it hard to explore the interplay of CDWs, SC, AHE, and nematicity. On the other hand, chemical doping via suitable partial element substitution can introduce carrier doping effect[28, 29], impurity effect[11], and chemical pressure[30], not only for the induction of SC but also serving as a 'probe' for the exploration of the SC mechanism. Specifically, cation substitution is one of the most effective ways for electron/hole doping in superconductors. For example, chemical substitutions in $CuO_2$ plane destroy the SC in the cuprate superconductors revealing the importance of short-range antiferromagnetic correlation of $Cu^{2+}$ ions[31], while some elements with the outermost electron configuration of $d^7$ and $d^8$ can induce SC by doping in the superconducting-active $Fe_2As_2$ or FeSe layer in Fe-based superconductors, indicating the relevance of SC to the itinerant $d$-electron and orbital fluctuation[32]. In kagome superconductor $LaRu_3Si_2$ with distorted kagome Ru layers, SC is suppressed due to a strong influence of impurity effect via the substitution of Ru by transition metal ions including Fe, Co, Ni and Cr[33], and the subsequent muon spin relaxation (μSR) experiment and calculation suggest SC is attributed to the kagome lattice related factors[34]. In a recent study, selective oxidation of exfoliated $CsV_3Sb_5$ thin flakes is reported to induce hole-doping to $CsV_3Sb_5$ and a superconducting dome was obtained as a function of doping content[35]. As the unusual properties are believed to come from the proximity of the V *3d* orbitals close to the van Hove singularity (VHS), chemical substitution of the transition metal V in the kagome layer should be an effective way to directly tune the VHS and unveil the origin of intrinsic electronic structures of $AV_3Sb_5$, but remains so far unexplored.

In this Letter, we report the synthesis of titanium-substituted $CsV_{3-x}Ti_xSb_5$ single crystals for the first time. We demonstrate that the nonmagnetic Ti atoms mainly substitute the V sites in the kagome layer through high-resolution scanning transmission electron microscopy (STEM) chemical imaging combined with density functional calculations. Low temperature scanning tunneling microscopy/spectroscopy (STM/S) reveals that both long-range $2a_0 \times 2a_0$ and $4a_0$ CDW and $4a_0/3 \times 4a_0/3$ PDW gradually disappear as the titanium content (labelled as *x*) increases. The magnetic susceptibility and transport measurements indicate that, as *x* increases, the onset temperatures of long-range CDW, nematicity, and AHE gradually decrease and are absent at around *x*=0.09, unequivocally demonstrating that the AHE and nematicity are strongly correlated to the long-range CDW. In addition, as *x* increases, $T_c$ firstly drops to a minimum value and then increases to 3.7 K, showing a double-dome shape in the phase diagram. Importantly, the Ti slightly substituted region shows an unconventional SC phase with a V-shaped gap, coexisting with long-range CDWs, nematicity, and AHE, whereas the Ti highly-substituted phase displays a conventional SC phase with a U-shaped gap coexisting with a short-range $4a_0$ CDW. These results demonstrate that the $CsV_{3-x}Ti_xSb_5$ exhibits two distinct SC phases with different correlated electronic states. The angle-resolved photoemission spectra (ARPES) measurements demonstrate that the Fermi level relatively shift to higher energy and the VHS at M point shift across the Fermi level as *x* increases, which may result in the suppression of CDW, AHE, and nematicity.

## 2. Materials and methods

2.1 *Single crystal growth of the Ti-substituted $CsV_3Sb_5$ samples.*

Single crystals of X-substituted (X=Ti, Cr and Nb) $CsV_3Sb_5$ were grown from Cs liquid (purity 99.98%), V powder (purity 99.9%), Ti, Cr or Nb shot (purity >99.9%) and Sb shot (purity 99.999%) *via* a modified self-flux method[36]. The mixture was put into an alumina crucible and sealed in a quartz ampoule under Argon atmosphere. The mixture was heated to 1000 ºC and soaked for 24 h, and subsequently cooled at 2 ºC/h. Finally, the single crystals were separated from the flux and the residual flux on the surface was carefully removed by Scotch tape. Except for sealing and heat treatment procedures, all other preparation procedures were carried out in an argon-filled glove box in order to avoid the introduction of any air and water. The obtained crystals have a typical hexagonal morphology with a maximum size of over 10×7×0.3 mm$^3$ (Fig. 1b) and are stable in air.

*2.2 Sample characterization.*

XRD patterns were collected using a Rigaku SmartLab SE X-ray diffractometer with Cu K$_\alpha$ radiation ($\lambda$ = 0.15406 nm) at room temperature. Scanning electron microscopy (SEM) and X-ray energy-dispersive spectroscopy (EDS) were performed using a HITACHI S5000 with an energy dispersive analysis system Bruker XFlash 6|60. Magnetic susceptibility and magnetization oscillation were determined by a SQUID magnetometer (Quantum Design MPMS XL-1 and MPMS-3). The SC transition of each sample was monitored down to 2 K under an external magnetic field of 1 Oe. Both in-plane electrical resistivity and Hall resistivity data were collected on a Quantum Design Physical Properties Measurement System (PPMS). Samples with different doping contents were dissolved by aqua regia for inductively coupled plasma (ICP) measurements.

*2.3 Scanning transmission electron microscopy and electron energy-loss spectroscopy*

Two cross-sectional samples along the [100] and [210] projections were prepared from the CsV$_{3-x}$Ti$_x$Sb$_5$ (*x*=0.15) single crystal using a focused ion beam (FIB) system. Atomic-scale STEM imaging and electron energy-loss spectroscopy (EELS) analyses were carried out on an aberration-corrected Nion U-HERMES100 dedicated STEM, operated at 60 kV. The probe forming angle was set to 32 mrad, and the probe current was set to 120 pA. The HAADF and annular bright field (ABF) images were collected from the 75-250 mrad and 15-30 mrad angular ranges, respectively. The EELS collection angle was set to 75 mrad. In order to reduce random noise in the EELS data, the STEM-EELS chemical mappings shown in the main text and SI have been denoised using principal component analysis (PCA). Quantification of the Ti:V ratio was calculated based on the integrated EELS signals of the two elements and the corresponding inelastic scattering cross-sections under the same experimental conditions. Note that this analysis is only semi-quantitative due to errors in calculating the inelastic scattering cross-sections for the different element edges.

*2.4 Scanning tunneling microscopy/spectroscopy*

The samples used in the experiments were cleaved at low temperature (about 13 K) and immediately transferred to an STM chamber. Experiments were performed in an ultrahigh vacuum ($1×10^{-10}$ mbar) ultra-low temperature STM system equipped with 11 T magnetic field. The electronic temperature in the low-temperature STS is 650 mK, calibrated using a standard superconductor, Nb crystal[37]. All the scanning parameters (setpoint voltage and current) of the STM topographic images are listed in the figure captions. Unless otherwise noted, the d$I$/d$V$ spectra were acquired by a standard lock-in amplifier at a modulation frequency of 973.1 Hz. The tungsten tips were fabricated via electrochemical etching and calibrated on a clean Au(111) surface prepared by repeated cycles of sputtering with argon ions and annealing at 500 °C. To eliminate the STM tip-drift effect from the topography and d$I$/d$V$ map, we apply the well-known Lawler-Fujita algorithm[38], after subtracting a 2$^{nd}$ or 3$^{rd}$ degree polynomial background, and obtain a set of displacement fields and drift-corrected topography. The so obtained displacement fields are then applied to the simultaneously measured d$I$/d$V$ maps in the same field of view as the topography.

*2.5 Density-functional theory (DFT) calculations*

Calculations are performed within the density-functional theory (DFT) as implemented in VASP package[39]. The generalized-gradient-approximation as parametrized by Perdew-Burke-Ernzerhof [40] for the exchange-correlation interaction between electrons is employed in all calculations. Zero damping DFT-D3 vdW correction[41] is also employed in all calculations while spin-orbital coupling is not included. The supercell method is used to calculate the formation energy. The supercells with Ti substitutions are fully relaxed until the remaining forces on the atoms are less than 0.005 eV/Å. The k-meshes of 6×6×3 and 3×3×4 are used to sample the Brillouin zones of the 2×2×2 and 3×3×1 supercell of $CsV_3Sb_5$, respectively. A cutoff energy of 300 eV for the plane-wave basis set is used.

*2.6 High resolution ARPES measurements*

High-resolution angle-resolved photoemission measurements were carried out on our lab system using ultraviolet laser as the light source that can provide a photon energy of $hv$=6.994 eV with a bandwidth of

0.26 meV. The energy resolution was set at ~2.5 meV for the measurements. The angular resolution is ~0.3°. The Fermi level is referenced by measuring on a clean polycrystalline gold that is electrically connected to the sample. The sample was cleaved in situ and measured in vacuum with a base pressure better than $5\times10^{-11}$ Torr.

## 3. Results and discussions

*3.1 Determination of the Ti substitution positions in $CsV_{3-x}Ti_xSb_5$ crystals*

We have successfully grown Ti-substituted $CsV_{3-x}Ti_xSb_5$ crystals with different substitution ratios ranging from $x=0.00$ to 0.27. The as-grown $CsV_{3-x}Ti_xSb_5$ single crystals (Methods) exhibit a stacking sequence of Cs-Sb2-VSb1-Sb2-Cs layers with hexagonal symmetry (space group P 6/mmm). In the VSb1 layer, the kagome sub-lattice of vanadium is interwoven with a simple hexagonal sub-lattice formed by the Sb1 atoms. As Ti is the neighboring nonmagnetic element to V, the ionic radii of $Ti^{4+}$ (60.5 pm) and $V^{5+}$ (54.0 pm) ion are similar, which results in a high possibility of Ti substitution at the V sites in the kagome lattice (Fig. 1**a**). A typical $CsV_{3-x}Ti_xSb_5$ crystal with a lateral size of over 1 cm and regular shape is shown in Fig. 1**b**. Smaller and thinner $CsV_{3-x}Ti_xSb_5$ single crystals showing regular hexagonal morphology (Fig. S1**a**), were also obtained. The Ti substitution content of a typical $CsV_{3-x}Ti_xSb_5$ single crystal determined by the energy-dispersive X-ray spectroscopy (EDS) is $x=0.15$ as shown in Fig. S1**b**, corresponding to 5% substitution of V. The representative X-ray diffraction (XRD) pattern and the rocking curve confirm the high-quality of the $CsV_{3-x}Ti_xSb_5$ single crystal with a preferred [001] orientation (Fig. S1**c** and **d**). The lattice parameters *a*, *b*, and *c* are measured to be 5.521, 5,521 and 9.336 Å, respectively, by single crystal diffraction, which are slightly smaller than those of pristine $CsV_3Sb_5$ (5.548, 5.548 and 9.349 Å, respectively). The lattice change is only about 0.4% and 0.1% in *a* and *c* axis, respectively, indicating a negligible chemical pressure.

To confirm the successful substitution of Ti into the $CsV_3Sb_5$ lattice, we carried out atomic-scale structural and chemical analysis on cross-sectional samples using aberration corrected STEM. Figure 1**c** shows a typical STEM high-angle annular dark-field (HAADF) Z-contrast image[42] of the $CsV_{3-x}Ti_xSb_5$ ($x=0.15$) sample along the [100] projection, with the structural models overlaid. The image clearly reveals the perfect crystalline structure of the $CsV_{3-x}Ti_xSb_5$ sample without noticeable structural defects, suggesting that Ti substitution does not degrade the crystal quality. The distinct sequential layers of Cs-Sb2-VSb1-Sb2-Cs are further confirmed by the elemental mapping shown in Fig. S2. Chemical analysis via electron energy-loss spectroscopy (EELS) unambiguously confirms the presence of Ti atoms (Fig. 1**d**), with the overall Ti substitution of V measured to be ~5 %, which is consistent with the EDS and inductively coupled plasma (ICP) measurements. The spatial distribution of the Ti atoms is revealed by the atomic-resolution chemical mapping shown in Figs. 1**e-g**. By comparing the simultaneously acquired V map (Fig. 1**e**) and Ti map (Fig. 1**f**), it is clear that most of the Ti atoms are located in the VSb1 layer, coinciding with the V atomic columns. Imaging and chemical analysis of the cross-sectional sample along the [210] projection (Fig. S3) reveal similar information. These results, thus, suggest that the majority of Ti atoms is substituting the V sites in the $CsV_3Sb_5$ lattice.

In order to understand the substitution mechanism, we estimated the formation energy of Ti substitution in $CsV_3Sb_5$ by density functional theory (DFT) calculations. We examined different Ti substitution contents and varied distances between neighboring Ti atoms. We found that the formation energy is in the range of –2.2 ~ –2.3 eV per Ti atom, which is insensitive to the Ti–Ti distance in the lattice (see more details in Fig. S4). The negative formation energies indicate that Ti atoms in $CsV_3Sb_5$ are energetically favored. The insensitivity of the formation energy to Ti-Ti distance suggests uniform distribution of Ti atoms inside $CsV_3Sb_5$, which is consistent with the observation by STEM.

*3.2 Evolution of correlated electronic states in $CsV_{3-x}Ti_xSb_5$ crystals*

We investigated the responses of CDW, AHE, and superconducting transitions to the substitution of Ti in $CsV_{3-x}Ti_xSb_5$ single crystals via the combination of magnetization, specific heat capacity, and electrical transport measurements. The superconductivity in $CsV_{3-x}Ti_xSb_5$ crystals with *x* between 0.00 and 0.27 was examined by diamagnetism and resistivity measurements. Bulk superconductivity with onset temperature $T_c^M$ is evidenced by the superconducting diamagnetism, as shown in Fig. 2**a**. It clearly shows that $T_c^M$ decreases to about 2.5 K at *x*=0.04 and then increases to 3.7 K at *x*=0.27. Importantly, the bulk superconductivity is confirmed by the 100% shield effect in $CsV_{3-x}Ti_xSb_5$ (*x*=0.04) based on the SC diamagnetism monitored down to 0.4 K (Fig. S5**a**). The onset superconducting transition temperature ($T_c^R$) derived from the temperature-dependent normalized resistivity of the pure $CsV_3Sb_5$ is above 4.0 K, about 0.5 K higher than that of the $CsV_{3-x}Ti_xSb_5$ (*x*=0.15) crystal (Fig. S5**b**). However, after applying different magnetic fields ranging from 0.02 to 1 T, the $T_c^R$ of the pure $CsV_3Sb_5$ shifts to lower temperatures with a broadening transition behavior, resembling the features of broad SC transitions observed in copper-oxide and iron-based high temperature superconductors[43-45], In contrast, the $T_c^R$ of the $CsV_{3-x}Ti_xSb_5$ (*x*=0.15) crystal decreases rapidly with applied fields showing a conventional superconductivity behavior similar to that of $MgB_2$[46].

To further investigate the Ti substitution effect on the normal state properties of this kagome superconductor system, systematic electrical transport measurements were conducted. Firstly, the CDW transition temperature ($T_{CDW}$) shows an obvious decease from ~94 K to ~60 K with increasing substitution contents, as evidenced by kinks in the temperature-dependent resistivity curves of $CsV_{3-x}Ti_xSb_5$ ($x$=0.00, 0.03, and 0.04) crystals (Fig. 2**b** and Fig. S5**c**). The kink fades out in $CsV_{3-x}Ti_xSb_5$ ($x$=0.09, 0.15, and 0.27) crystals, indicating the absence of long-range CDWs. The temperature-dependent specific heat-capacity measurements (Fig. S5**d**) also provide strong evidences for the undetectability of long-range CDWs in $CsV_{3-x}Ti_xSb_5$ ($x$=0.09, 0.15, and 0.27) crystals.

We performed Hall measurements on $CsV_{3-x}Ti_xSb_5$ crystals with different substitution contents at various temperatures (Fig. S6**a,b,c,d,e**). $\rho_{xy}$ of the $CsV_{3-x}Ti_xSb_5$ crystal ($x$=0.03) exhibits a linear behavior at high temperature and an antisymmetric sideways "S" line shape at low fields below $T_{CDW}$, which is similar to those of pure $CsV_3Sb_5$ due to the AHE. In contrast, $\rho_{xy}$ of $CsV_{3-x}Ti_xSb_5$ crystals ($x$=0.04, 0.09, and 0.15) show a linear behavior over the entire range of external fields under different temperatures. Unlike a significant enhancement of hole mobility of pure $CsV_3Sb_5$ at low temperature as a multiband system[9], Hall coefficients $R_H$ of $CsV_{3-x}Ti_xSb_5$ crystals are negative within a wide temperature range from 150 K down to 2 K, indicating that electron-type carriers are dominant (Fig. S6**f**). A gradual increase of the absolute value $R_H$ can be observed upon substitution contents. $\rho_{xy}^{AHE}$ of $CsV_{3-x}Ti_xSb_5$ crystals ($x$=0.03 and 0.04) persists up to about 60 K (Fig. S6**g,h**) and $\rho_{xy}^{AHE}$ of $CsV_{3-x}Ti_xSb_5$ crystals ($x$=0.09) demonstrates no AHE. The anomalous Hall conductivity (AHC) $\sigma_{xy}^{AHE}$ of $CsV_{3-x}Ti_xSb_5$ crystals ($x$=0.03 and 0.04) clearly shows the sudden appearance of the AHE below $T_{CDW}$ and the values of $\sigma_{xy}^{AHE}$ decreases with the Ti substitution increasing (Fig. 2**c**), which is possibly due to the suppression of the long-range CDWs.

We also measured the in-plane angular-dependent magnetoresistance (AMR) of $CsV_{3-x}Ti_xSb_5$ ($x$=0.03, 0.04, 0.09, and 0.15) crystals under a field of 5 T at different temperatures above $T_c$. The ratios of AMR defined as $\Delta R/R_{min} = [R(\theta,T)-R_{min}(T)]/R_{min}(T)\times100\%$ are summarized in Fig. S7 by polar-coordinate plots. AMR of $CsV_{3-x}Ti_xSb_5$ ($x$=0.00, 0.03, 0.04) crystals appears below $T_{CDW}$ and the ratios of AMR decrease with the increasing Ti substitution content. The $CsV_{3-x}Ti_xSb_5$ ($x$=0.09 and 0.15) crystals show no AMR signals in their normal states (Fig. S7**f**). The ratios of AMR at 5 K indicates clearly the absence of the two-fold rotational symmetry with high Ti substitution content of $x$=0.09 (Fig. 2**d**), namely the suppress of nematicity by Ti substitution[47].

To understand the evolution of CDW, AHE, and AMR in CsV$_{3-x}$Ti$_x$Sb$_5$ crystals, the band structures crossing Brillioun zone center Γ along Γ-K direction for different Ti substitution contents of x=0.00, 0.04, 0.15, and 0.27 have been performed at 20 K by high-resolution VUV laser ARPES (Fig. S8). An electron like band around Γ point (α band) is present for all the Ti-substituted samples. At $x$=0.00, some extra bands around Γ just below the Fermi level (E$_F$) and around the binding energies between 0.4 and 0.8 eV can be observed. These bands are from the band folding of those around M points due to the 2×2 reconstruction in the CDW state[48]. They are absent for the CsV$_{3-x}$Ti$_x$Sb$_5$ samples, which is consistent with our results that the long-range CDW state is suppressed by the Ti substitution (Fig. S8**a**). The energy distribution curves (EDCs) at Γ point (Fig. S8**b)** show the energy positions shift up monotonously by about 5, 10 and 100 meV for the CsV$_{3-x}$Ti$_x$Sb$_5$ samples ($x$= 0.04, 0.15 and 0.27) as compared to that of the pure sample. The shift of 100 meV for CsV$_{3-x}$Ti$_x$Sb$_5$ samples ($x$= 0.27) clearly indicates that the hole doping pushes the Fermi level downwards. The DFT calculations also demonstrate that with increasing Ti substitution the Fermi energy move down with slight dispersion changes (Fig. S8c,d). The above ARPES and DFT results indicate that the substitution of Ti is an efficient way to tune the Fermi level of the CsV$_3$Sb$_5$ system by hole doping.

Comparing the band structures around M point and *Γ* point between the CsV$_3$Sb$_5$ and CsV$_{3-x}$Ti$_x$Sb$_5$ (x=0.27) samples (Fig. S9), the bands just below E$_F$ around M (marked by pink and purple lines in Fig. S9**c**) in the pure CsV$_3$Sb$_5$, representing the VHS2 bands marked in Fig. S8**d**, are absent (pink) or crossing the E$_F$ (purple) around M point in the CsV$_{3-x}$Ti$_x$Sb$_5$ ($x$=0.27) sample (Fig. S9**d**). The ARPES results clearly indicate that the VHS2 of electronic structures at *M* point shifts across over the Fermi level after Ti substitution in the kagome lattice. Therefore, we conclude that the suppression of electron scattering around *M* points, and the corresponding suppression of CDW, originates from the crossing of VHS2 point above E$_F$ at M point. The suppression of CDW results in the suppress of other correlated electronic states including AHE, AMR, and nematicity.

*3.3 Local electronic structures of CsV$_{3-x}$Ti$_x$Sb$_5$ crystals*

We then studied the evolution of CDW on the surface of the CsV$_{3-x}$Ti$_x$Sb$_5$ samples with various Ti substitution contents at the atomic scale by low temperature STM/S (Fig. 3). In the STM measurements, almost all the surface regions show large-scale Cs surface topography. Large-scale Sb surface topography was rarely observed. Because of the reconstruction of Cs terminated surface (Fig. S10) and its possible surface doping effect[49, 50], we choose Sb surface to study the evolution of CDWs. We applied the STM manipulation method to sweep the top Cs atoms away to expose large-scale Sb surfaces[6]. In both STM topography (T(**r**,V)) and d$I$/d$V$ maps (d$I$/d$V$(**r**,V)) of the large-scale Sb surface, we found that, as compared to the pure CsV$_3$Sb$_5$ sample[6], many small spots with darker intensity appear and randomly distribute in the STM images of the Sb surface. Atomically resolved STM images demonstrate that the Sb lattice remains continuous across the dark spots (inset in Fig. 3**a**), indicating that the dark spots originate from electron scattering in the underlying VSb1 layer. We then analyzed the height histogram of the STM images of the Sb surface and found that the area density of the dark spots shows a positive correlation with the Ti substitution contents (Fig. S11), which further qualitatively demonstrates that the dark spots in the STM images originate from Ti atoms in the underlying VSb1 kagome layer.

In addition to the emergence of dark spots, there are significant changes of the CDW in both STM images and d$I$/d$V$ mapping of the CsV$_{3-x}$Ti$_x$Sb$_5$ samples as compared with the pure CsV$_3$Sb$_5$ sample. We find that the bidirectional 2$a_0$ × 2$a_0$ (green circles in Fig. 3**c**) and unidirectional 4$a_0$ CDWs (red squares in Fig. 3**c**) at all energies (details in Fig. S12) are preserved for the slightly-substituted CsV$_{3-x}$Ti$_x$Sb$_5$ samples ($x$=0.03 in Figs. 3**a-c** and $x$=0.04 in Fig. 3**d-f**). However, both 2$a_0$×2$a_0$ and 4$a_0$ CDWs almost disappear in the Fourier transform of the topography and d$I$/d$V$ mapping of the $x$=0.15 sample (Figs. 3**g-i**).

The simultaneous suppression and disappearance of long-range 2$a_0$×2$a_0$ and 4$a_0$ charge orders strongly indicate that these two CDWs are intertwined with each other. It should be noticed that, the bidirectional 4/3$a_0$ PDW peaks[6] (labeled by pink circles in Fig. S12**a,b**) in the Fourier transform of d$I$/d$V$ maps at low energy (< 5 meV) are suppressed in the $x$=0.03 and 0.04 samples but invisible in the $x$=0.15 sample (Fig. S12**c**), suggesting that the PDW is intertwined with 2$a_0$×2$a_0$ and 4$a_0$ CDWs.

Although the peaks correspond to the long-range 2$a_0$×2$a_0$ and 4$a_0$ CDWs in the FT are too weak to be observed, short-range striped orders can be clearly seen in the topography of the CsV$_{3-x}$Ti$_x$Sb$_5$ ($x$=0.15) sample (Fig. 3**j**). Furthermore, the temperature-dependent STM images of the Sb surface of CsV$_{3-x}$Ti$_x$Sb$_5$ ($x$=0.15) show that the short-range stripe orders disappear at a critical temperature of ~55 K (Fig. 3**j**), similar to the critical temperature in the pure CsV$_3$Sb$_5$ [51].

*3.4 Microscopic evolution of superconductivity and phase diagram of CsV$_{3-x}$Ti$_x$Sb$_5$ crystals*

To reveal the microscopic evolution of superconductivity in $CsV_{3-x}Ti_xSb_5$ crystals, we lowered the electron temperature to 650 mK (Fig. S13) and collected a series of d$I$/d$V$ spectra at both Sb and Cs surfaces (Fig. S10). Because the superconducting gap size is found to be homogenous despite of some variations in the coherence peak height and gap depth in different regions of sample surface (Fig. S14), we use the spatially-averaged d$I$/d$V$ spectra over tens of regions for comparison. We find that the $x$=0.03 sample shows a V-shaped paring gap (blue curve in Fig. 4a) corresponding to a nodal superconducting phase, whereas the $x$=0.15 and 0.27 samples show a conventional Bardeen–Cooper–Schrieffer (BCS) gap (details please see Fig. S15 and S16) corresponding to a nodeless superconducting phase (green and dark green curve in Fig. 4a). The V-shaped pairing gap in the $x$=0.03 sample is slightly smaller than that of the pure sample (black curve), and the zero-bias conductance is much higher than that in the pure sample, indicating the decrease of $T_c$. For the $x$=0.15 and 0.27 samples, the SC paring is totally different from the pure and slightly substituted samples, demonstrating the transition into a new SC phase.

Figure 4b presents the phase diagram of $CsV_{3-x}Ti_xSb_5$ single crystals, where the onset temperatures of long-range three-dimensional CDW ($T_{CDW}$) and superconductivity ($T_c$), together with the amplitude of AHE and AMR (top panel), are summarized as a function of substitution content $x$. Both $T_{CDW}$ and $T_c$ are suppressed with increasing Ti substitution and $T_{CDW}$ become absent below $x$=0.09, accompanied by the strongly correlated evolution of the AHE and AMR amplitude (Fig. 4b upper panel), clearly demonstrating that they are strongly correlated with each other. In addition, the $CsV_{3-x}Ti_xSb_5$ system shows two superconducting phases. In the Ti slightly-substituted phase, the superconducting state show a V-shape pairing gap coexisting with long-range charge orders, AHE and AMR. The $T_c$ decreases from ~3.5 K of the pure $CsV_3Sb_5$ sample to ~2.0 K of the $x$=0.04 sample. In the Ti highly-substituted phase, the $T_c$ increases from ~2.0 K ($x$=0.04) to ~3.7 K ($x$=0.27). This superconducting phase exhibits the U-shaped gap pairing and a sharp SC transition at the expense of $T_c$ under magnetic field (Fig. S5b, solid lines). Although the long-range charge orders are undetectable in the second phase, there are still short-range stripe orders, which indicates that the coexistence of correlated electronic states and U-shaped superconductivity featuring parallels to iron-based superconductors. To our knowledge, $CsV_{3-x}Ti_xSb_5$ is the first kagome superconductor reported so far that two distinct SC phases emerge upon chemical substitution at ambient pressure, strongly correlating with CDW, AHE, and AMR. Such a transition from V-shaped to U-shaped superconducting phases is rare in the study of superconductivity, and is in analogous to the twisted trilayer graphene [52].

It is worth to note that the V kagome layer substitution is not limited to Ti. We further explored the substitution of V with elements in the same periodic or the same group elements of V. Using the same synthesis method, we have successfully prepared $CsV_{3-x}Cr_xSb_5$ ($x$=0.06 and 0.15) and $CsV_{3-x}Nb_xSb_5$ ($x$=0.03 and 0.06) crystals through substitution of V ions in the kagome layer with Cr ($Cr^{5+}$~49.0 pm) and Nb ($Nb^{4+}$~68.0 pm). For the Cr-substituted samples, the SC is suppressed dramatically as Cr substitution content increases and becomes undetectable at $x$=0.15, and the CDW features become significantly weaker (Fig. S17). In contrast, for the Nb-substituted samples, the SC is enhanced ($T_c$ is 4.2 K for $x$=0.06, which is higher than 3.5 K of $CsV_3Sb_5$) and CDW signals fade gradually as Nb substitution content increases (Fig. S18). Therefore, the present work opens up a way to synthesise a new family of doped kagome materials by various substitution elements, and represents a new platform for tuning the correlated electronic states including charge orders, loop current, nematicity and superconducting pairing.

## 4. Summary

We report, for the first time, the synthesis of Ti-doped vanadium-based kagome superconductor $CsV_{3-x}Ti_xSb_5$. STEM demonstrate that the V atoms in the kagome layers is directed substituted by the Ti atoms. $CsV_{3-x}Ti_xSb_5$ exhibits two distinct SC phases upon substitution. The Ti slightly-substituted phase displays an unconventional V-shaped SC gap, coexisting with weakening CDW, PDW, AHE, and nematicity. The Ti highly-substituted phase has a U-shaped SC gap concomitant with a short-range rotation symmetry breaking CDW, while long-range CDW, twofold symmetry of in-plane resistivity, AHE, and PDW are absent. The high-quality single crystals with controlled Ti substitution contents in kagome planes provides a new platform allowing for systematic study of the evolution of superconductivity with multiple intertwining orders (CDW, PDW, AHE, and nematicity) in both macroscopic and microscopic views, which is not possible with high-pressure approaches and/or polycrystalline samples. The double-dome superconductivity only indicates the change of $T_c$, whereas the microscopic evolution of SC pairing upon chemical doping and pressure is more important for understanding the kagome-lattice based physics. We discover a unique phase exhibiting U-shaped SC gap concomitant with short-range rotation-symmetry breaking orders in the highly-substituted crystals, featuring parallels to iron-based superconductors. Such a transition from V-shaped to U-shaped transition superconducting phases concomitant with the evolution of the long-rang to short-rang CDW provides a rare platform in the study of superconductivity. The systematic and detailed mechanism of the competing orders in the Ti-doped $CsV_3Sb_5$ by both experiments and theory need further studies in the future.

**Note**: During the reviewing process, it came to our attention that double-dome superconductivity has been recently reported in CsV$_3$Sb$_{5-x}$Sn$_x$ polycrystalline powder samples[53].

**Conflict of interest**

The author declare that they have no conflict of interests.

**Acknowledgements**

We thank Zhong-Xian Zhao and Gang Su for helpful discussions. The work is supported by grants from the National Natural Science Foundation of China (61888102, 52022105, 51771224, 11888101, 12061131005 and 11834016), the National Key Research and Development Projects of China (2018YFA0305800 and 2019YFA0308500), the Chinese Academy of Sciences (XDB33030100, XDB28010200, and XDB30010000), the Key Research Program of Chinese Academy of Sciences (ZDBS-SSW-WHC001), the CAS Project for Young Scientists in Basic Research (YSBR-003) and the Beijing Outstanding Young Scientist Program (BJJWZYJH01201914430039). Z.W. is supported by the US DOE, Basic Energy Sciences Grant No. DE-FG02-99ER45747. B.Y. acknowledges the financial support by the European Research Council (ERC Consolidator Grant, No. 815869) and the Israel Science Foundation (ISF No. 1251/19).

**Author Contributions**

Hong-Jun Gao supervised and coordinated the project. Hong-Jun Gao and Haitao Yang designed the research. Haitao Yang and Zhao Zhen prepared the samples. Yuhang Zhang, Zouyouwei Lu, Hua Zhang, Jianping Sun, Jingguang Cheng, Haitao Yang and Xiaoli Dong performed the transport experiments. Hui Chen, Zihao Huang, Guojian Qian, and Bin Hu performed the STM experiments with guidance of Hong-Jun Gao. Wu Zhou, Jinan Shi and Stephen J. Pennycook performed the STEM experiments. Hailan Luo, Lin Zhao and Xingjiang Zhou perform the ARPES experiments. Tengxin Tan and Binghai Yan performed the DFT calculations. All authors participated in the data analysis and manuscript writing.

**Main Figures**

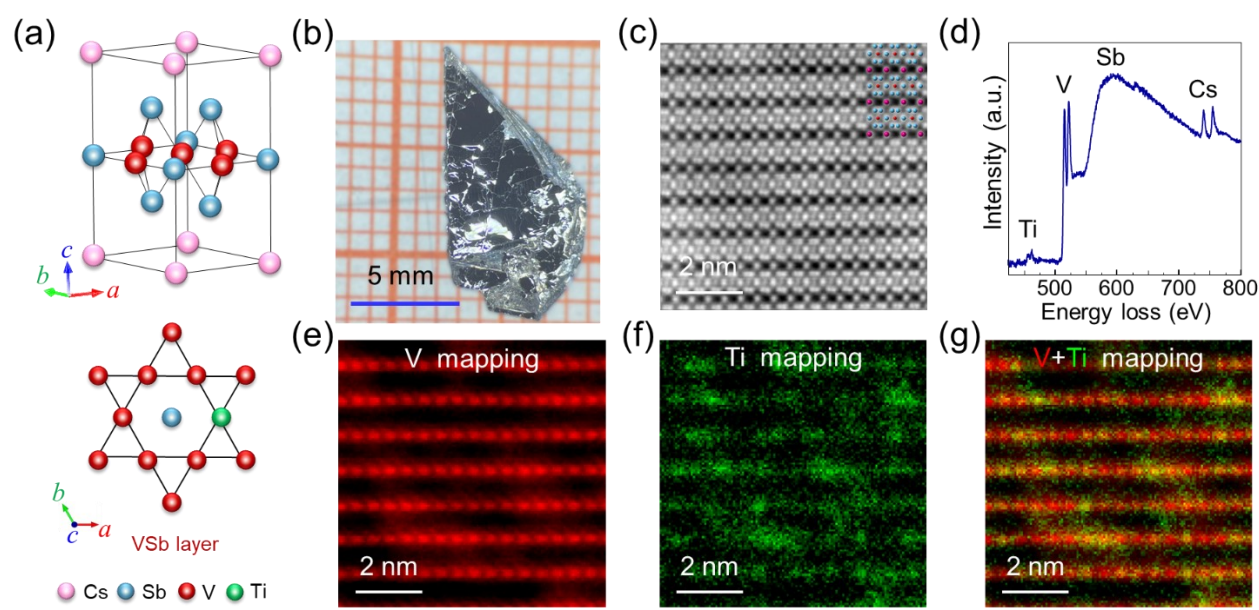

**Fig. 1.** Determination of the Ti substitution positions in $CsV_{3-x}Ti_xSb_5$ single crystals. (a) Schematic of atomic structure of Ti-substituted $CsV_{3-x}Ti_xSb_5$ crystal with Cs atoms in light purple, Sb atoms in light blue, V atoms in red and Ti atoms in green. The Ti atoms replace the V atoms in the kagome lattice. (b) A photo of a single crystal of the as-prepared $CsV_{3-x}Ti_xSb_5$ ($x$=0.15) crystal. (c) STEM-HAADF Z-contrast image of $CsV_{3-x}Ti_xSb_5$ crystal viewed along the [100] projection, with the atomic structural models overlaid. The Cs, V, Sb atoms are shown in purple, red, and light blue, respectively. (d) Background-subtracted EELS spectrum showing clear Ti, V, Sb, and Cs signals. Quantification of the Ti and V signals shows an atomic ratio of Ti:V ~5.3:94.7. (e-f) Atomic-resolution chemical mapping acquired *via* STEM-EELS spectrum imaging, with the simultaneously acquired V mapping shown in red (e) and Ti mapping shown in green (f). g, Overlay of the V (red) and Ti (green) signals. Ti is mostly doped into the V sites.

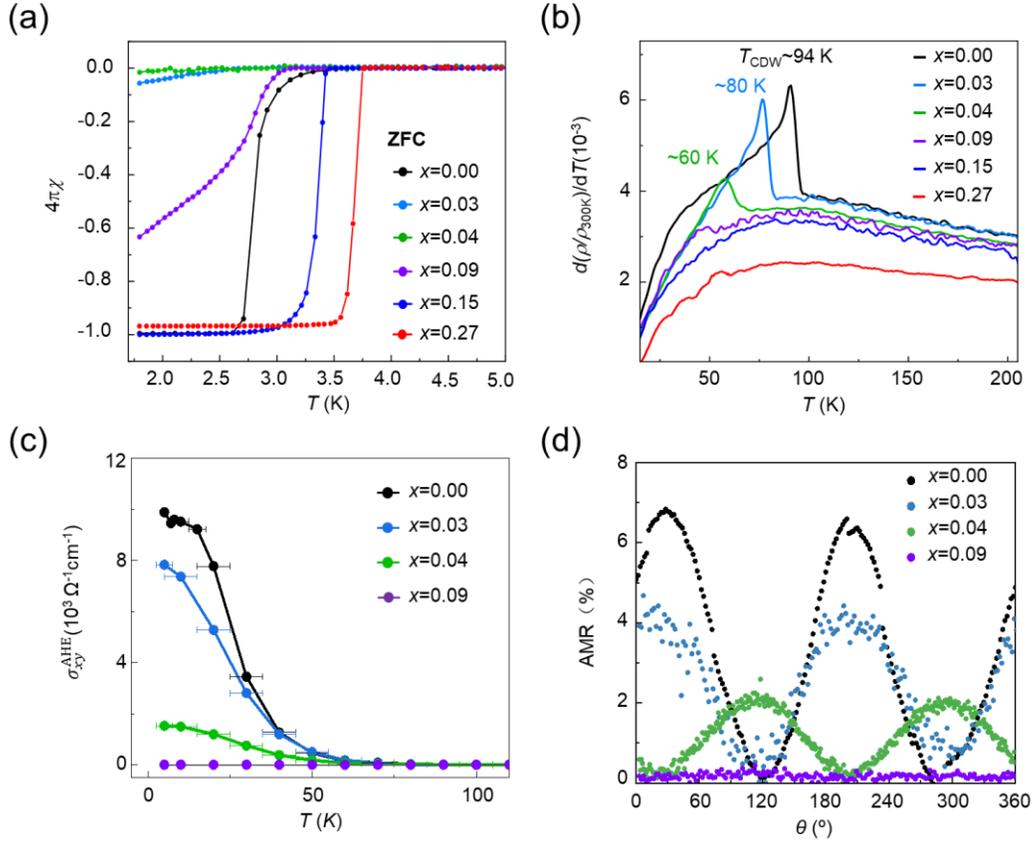

**Fig. 2.** Ti substitution effects on SC, CDW, AHE, and AMR in CsV$_{3-x}$Ti$_x$Sb$_5$ crystals. (a) The temperature-dependent magnetic susceptibilities corrected for demagnetization factor under zero-field cooling (ZFC) with applied field of 1 Oe along the $c$ axis. (b) The derivative electrical resistivity d$\rho$/d$T$ curve of CsV$_{3-x}$Ti$_x$Sb$_5$ crystals with $x$ from 0.00 to 0.27. The CDW is suppressed upon increasing $x$ and undetectable when $x$ exceeds 0.09. (c) The anomalous Hall conductivity of CsV$_{3-x}$Ti$_x$Sb$_5$ ($x$=0.00, 0.03, 0.04, and 0.09) crystals as a function of temperature obtained by subtracting the local linear ordinary Hall conductivity background. (d) The in-plane angular-dependent magnetoresistivity (AMR) equated to $\Delta R/R_{min}$ as a function of $\theta$ at $T$=5 K. Here $\theta$ is the angle between the directions of the external field ($H$) of 5 T and the current ($I$), with $\theta$ = 0° corresponding to $H$ // $I$.

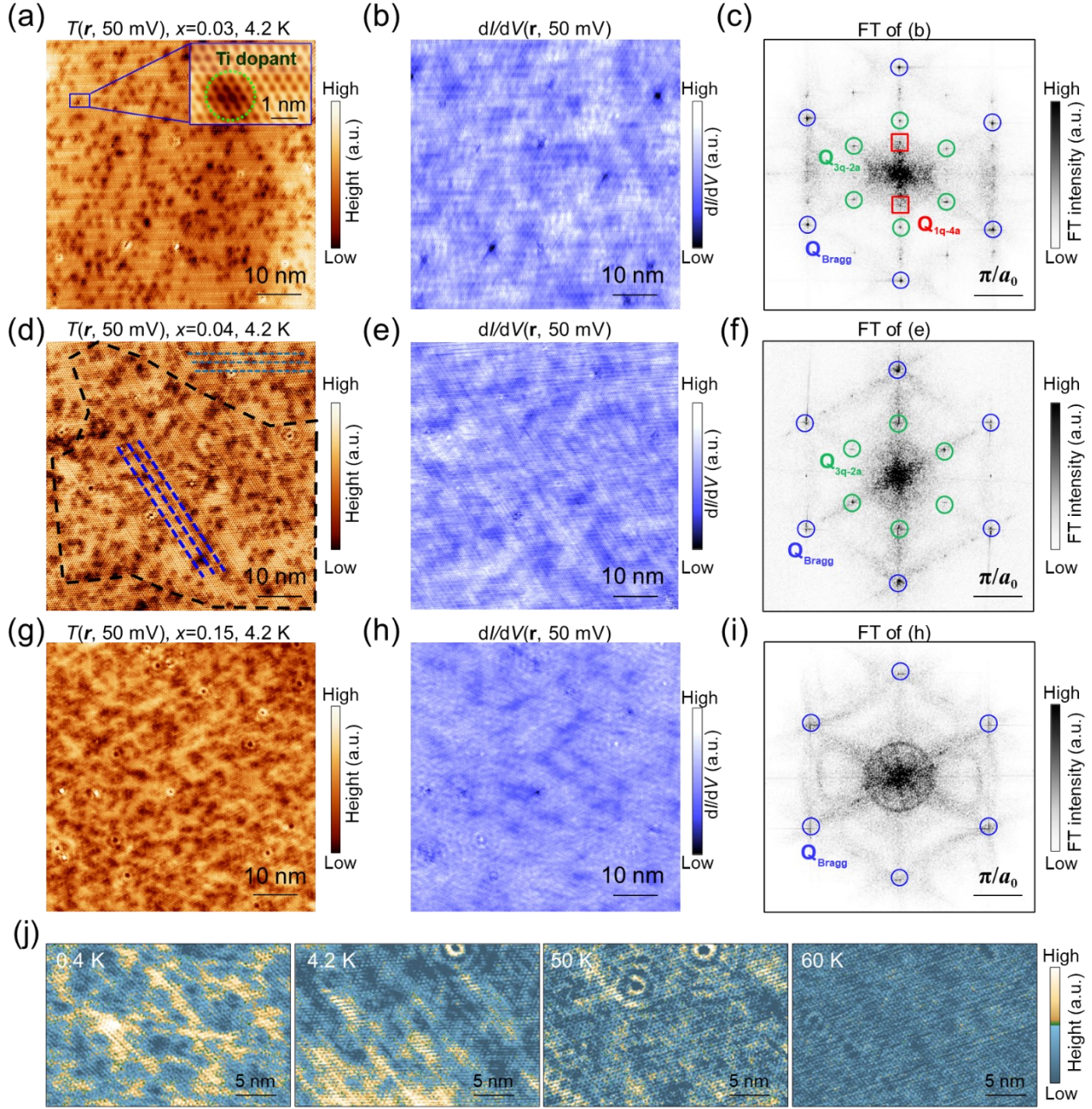

**Fig. 3.** Local electronic structures of CsV$_{3-x}$Ti$_x$Sb$_5$ crystals. (a-c) STM topography (a), d$I$/d$V$(**r**, 50 mV) (b) and the corresponding drift-corrected Fourier transform (FT) (c) at the large-scale Sb surface of CsV$_{3-x}$Ti$_x$Sb$_5$ ($x$=0.03) obtained at 4.2 K, showing the 3Q-2$a$ and 1Q-4$a$ CDWs. Many dark spots, correlated to the Ti atoms, appear in the STM image of (a). The Inset in (a): Zoom-in STM image showing the continuous lattice around the dark spots. (d-f) STM topography (d), d$I$/d$V$(**r**,-5 mV) (e) and the corresponding drift-corrected FT (f) at the large-scale Sb surface of CsV$_{3-x}$Ti$_x$Sb$_5$ ($x$=0.04), showing the presence of short-range 1Q-4$a$ CDW multiple domains. Two 4$a$ CDW domains can be clearly observed (highlighted by blue and light blue dotted lines). (g-i) STM topography (g), d$I$/d$V$(**r**, 50 mV) (h) and the corresponding drift-corrected FT (i) at the large-sized Sb surface of $CsV_{3-x}Ti_xSb_5$ ($x$=0.15), showing the absence of 3Q-2$a$, and long-range 1Q-4$a$ CDW states in the FT. The weak and elongated stripes around M points in (i) result from QPI pattern instead of long-range 3Q-2$a$ CDWs. Setting parameter: tunneling current setpoint $I_t$=1.0 nA. (j) STM images of the Sb surface of $CsV_{3-x}Ti_xSb_5$ ($x$=0.15) at 0.4 K, 4.2 K, 50 K and 60 K, respectively, showing that the short-range charge orders disappear at a critical temperature of ~60 K. $V_s$=-50 mV, $I_t$=1.0 nA.

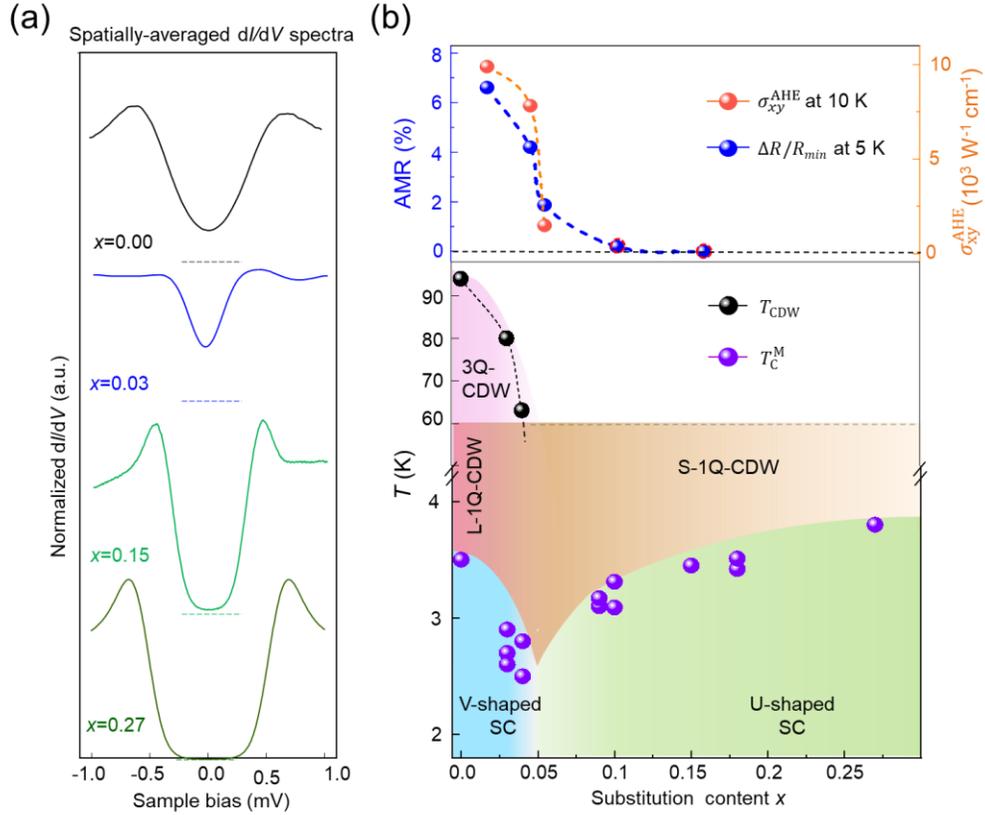

**Fig. 4.** Phase diagram of $CsV_{3-x}Ti_xSb_5$ crystals. (a) Spatially-averaged $dI/dV$ spectra obtained on the Sb surfaces of the $CsV_3Sb_5$ (black curve) and $CsV_{3-x}Ti_xSb_5$ samples ($x$=0.03, 0.04, 0.15 and 0.27, corresponding to blue, green, and dark green curves, respectively), showing a transition from V-shape to U-shape symmetry through Ti substitution. The horizontal dash lines highlight the positions of zero density of states for each curve. (b) Phase diagram of the $CsV_{3-x}Ti_xSb_5$ crystals. In the V-Shaped SC regime, $T_{CDW}$ is significantly reduced with increasing $x$. The amplitude of the anomalous Hall conductivity $\sigma_{xy}^{AHE}$ and AMR (%), is dramatically decreased, that is, long-range CDW, AHE and nematicity were simultaneously suppressed. In contrast, in the U-shaped SC regime where AHE, long-range CDW, and nematicity are undetectable, the superconductivity coexisting with short-range CDWs tends to gradually recover: $T_c^M$ rises to ~3.5 K at $x$= 0.15, and showing a slight increase to 3.7 K at $x$=0.27.